\documentstyle[preprint,amsfonts,aps]{revtex}

\begin{document}
\draft
\title{The liquid-vapor interface of an ionic fluid}
\author{B. Groh$^1$, R. Evans$^{1,2}$ and S. Dietrich$^{1}$}
\address{
 $^2$Fachbereich Physik, Bergische Universit\"at Wuppertal, \\
 D--42097 Wuppertal, Federal Republic of Germany\\
{$^2$H.H. Wills Physics
    Laboratory, University of Bristol, \\Bristol BS81TL, U.K.}}
\date{\today}

\maketitle

\begin{abstract}

We investigate the liquid-vapor interface of the restricted primitive
model (RPM) for an ionic fluid using a density-functional
approximation based on correlation functions of the homogeneous fluid
as obtained from the mean-spherical approximation (MSA). In the limit
of a homogeneous fluid our approach yields the well-known MSA (energy)
equation of state. The ionic interfacial density profiles, which for
the RPM are identical for both species, have a shape similar to those
of simple atomic fluids in that the decay towards the bulk values is
more rapid on the vapor side than on the liquid side. This is the
opposite asymmetry of the decay to that found in earlier calculations
for the RPM based on a square-gradient theory. The width of the
interface is, for a wide range of temperatures, approximately four
times the second moment correlation length of the liquid phase. We
discuss the magnitude and temperature dependence of the surface
tension, and argue that for temperatures near the triple point the
ratio of the dimensionless surface tension and critical temperature is
much smaller for the RPM than for simple atomic fluids.
\end{abstract}
\pacs{PACS numbers: 68.10.-m, 61.20.Qg}

\section{Introduction}

In this paper we develop a density-functional theory (DFT) for the
properties of the liquid-vapor interface of the simplest model of an
ionic fluid, namely the restricted primitive model (RPM) in which the
ions are modelled by equisized hard spheres of equal and opposite
charge. The RPM serves as a simple model for molten salts and
electrolyte solutions. Indeed the measured partial structure factors
$S_{ij}(k)$, with $i,j\in\{+,-\}$, and the thermodynamic properties of
several molten alkali halides near their melting points can be well
described by the corresponding quantities of the RPM
\cite{Parrinello}. In recent years there has been a revival of
interest in the properties of the RPM, stemming from efforts to
understand the nature of criticality in ionic fluids
\cite{FisherJSP,FisherNorwich,CritExp}. For ionic systems one might
suppose that the long-ranged Coulomb forces could give rise to
critical exponents different from the Ising ones which are measured
and calculated for atomic and molecular fluids. Some recent
experiments on certain electrolytes revealed mean-fieldlike
behavior or, in some cases, Ising critical regions which are several
orders of magnitude smaller than in atomic fluids
\cite{CritExp}. Since the three-dimensional RPM is known to exhibit
phase separation into a dense, conducting ionic liquid and a very
dilute vapor phase, which is also conducting, it is a natural choice
for theoretical and simulation studies of phase coexistence and
criticality. Although attempts to explain the experimental observations
regarding criticality have so far been unconvincing --- estimates of
the Ginzburg temperatures for the RPM are similar to those for simple
(atomic) fluids \cite{Raul2,FisherPRL} and the latest Monte Carlo
finite size scaling study \cite{CaillolRPMMC} gives results compatible
with Ising behavior --- this does not mean that the RPM does not
warrant further attention. On the contrary, because it incorporates
the key features of hard core repulsion and Coulomb forces it remains
the canonical, albeit over-idealized, model for an ionic fluid.

Here we shift attention away from the bulk and focus on the
inhomogeneous situation which arises at the planar interface between
the coexisting liquid and vapor of the RPM. We are interested in the
ionic density profiles and the surface tension of such an interface,
which can be viewed as a crude model for the corresponding
liquid-vapor interface of a molten alkali-halide. Compared with the
well studied case of simple fluids, modelled by a Lennard-Jones
potential \cite{RW}, very little is known about the interface in the
RPM. We are not aware of any corresponding simulation studies,
although there is one early molecular dynamics simulation \cite{Heyes} using
the more realistic Born-Mayer-Huggins potential model for
KCl. 

Theoretical work was pioneered by Telo da Gama et al. \cite{GT}
who used a gradient expansion developed in Ref.~\cite{EvSluck} to
investigate the RPM interface. The special symmetry of the RPM implies
that the density profile of the cations should be the same as that of
the anions, i.e., there should be local electroneutrality
$\rho_+(z)=\rho_-(z)$ throughout the interface. For such a
``symmetric'' situation the gradient expansion of the free energy
functional involves only gradients of the total density profile
$\rho(z)=\rho_+(z)+\rho_-(z)$; the charge density profile
$q(z)=\rho_+(z)-\rho_-(z)$ vanishes identically. The coefficients in
this expansion involve moments of the density-density bulk direct
correlation function $c_\rho^{(b)}(r;\rho)$. If the expansion is
truncated at the square gradient term, as is usual, the corresponding
coefficient must be positive if the theory is to yield physical
solutions. As was explained in Ref.~\cite{GT}, it is necessary to utilize a
rather sophisticated, self-consistent theory of the bulk correlation
functions, the generalized mean-spherical approximation (GMSA)
\cite{HoyeGMSA}, in order to obtain a positive coefficient. The
ordinary MSA \cite{Waisman1,Waisman2}, which is often successfully
employed \cite{Parrinello}
in studies of the bulk RPM, is insufficient as it yields merely the hard
sphere (Percus-Yevick) result for $c_\rho^{(b)}(r;\rho)$ which
produces a {\it negative} coefficient of the square gradient
term. That one must supercede the MSA to obtain physical interfaces
is, at first sight, quite surprising --- given its success in
bulk. One suspects that this is an artefact of the square-gradient
approximation. Here we re-investigate the liquid-vapor interface using
an alternative DFT, which does not utilize the gradient expansion and
thereby avoids the need to employ the GMSA and the ensuing problems
involved with extrapolation into the two-phase region \cite{GT}. Our
approach, which is motivated by the MSA treatment of the bulk free
energy, involves a local density approximation for the hard-sphere part
of the functional and a non-local treatment of the remaining
(Coulombic) contributions, which is obtained by approximating the
inhomogeneous pair correlation functions by their homogeneous
counterparts. It differs from other DFT approaches for ionic fluids
\cite{Mier,Groot,Kierlik,Ghosh,AttardRev} which have proved successful
in the primitive model description of electrical double layers at
charged hard walls. These theories treat the non-hard-sphere part of
the functional by means of a second order density expansion about the
density of a reference fluid, usually taken to be the homogeneous bulk
fluid far from the substrate. Although this is adequate for many
purposes it is problematical when it comes to liquid-vapor interfaces
or to the adsorption of thick (wetting) films, where {\it two} bulk
phases are involved. Indeed, for the case of an atomic fluid the
corresponding second order expansion about a homogeneous reference
density is known to fail to account for liquid-vapor coexistence and
is inadequate for wetting problems \cite{Evans83}. We are not aware of
attempts to use the approaches in Refs.~\cite{Mier,Groot,Kierlik,Ghosh,AttardRev} for the liquid-vapor
interface. Nor are we aware of attempts to use integral-equation
theories for that purpose which have been rather popular in studies of the electrical
double layer \cite{BlumRev,AttardRev} but which may be beset by
similar problems.

The present approach does not suffer from these difficulties, i.e., the
uniform limit of the free energy reduces to that of the MSA for any
uniform density $\rho_b$. On the negative side this means that our
theory for the interface is prone to the same deficiencies as the bulk
MSA, namely the failure to incorporate properly the effects of ion
pairing, which are especially pronounced in the vapor, and the
consequent poor estimate of the location of the critical point.

As well as providing a description of the liquid-vapor interface of a
near-symmetric alkali halide, i.e., one where the ions have nearly the
same diameter, the present theory may form the basis for a description
of wetting phenomena in ionic fluids \cite{Kayser}. The situations one would like to
consider are: (i) the wetting of a substrate-alkali halide vapor
interface by the molten salt, (ii) the wetting of the interface
between a substrate and, say, the phase dilute in salt by the other,
salt-rich, phase for an electrolyte that exhibits liquid-liquid phase
separation such as those considered in Refs.~\cite{CritExp} and \cite{ionicwet}. Since it is well known
that wetting properties depend sensitively on the range of the
interaction potentials, one might ask if the long-range Coulomb
interactions lead to qualitatively new features or if, due to
screening, ionic fluids behave as one-component systems
with short range interactions, e.g., Yukawa fluids.

This paper is arranged as follows. In Sec.~\ref{SecDFT} we introduce
the approximate DFT for our particular problem. Section~\ref{Secbulk}
examines the free energy and the second moment correlation length which
emerge for the {\it bulk} fluid, comparing and contrasting the results
with those of other theories and with exact results for the limit
$\rho\to 0$. In Sec.~\ref{Secinterf} we describe the results for the
density profile and surface tension of the liquid-vapor interface as a
function of temperature. These are compared with those of the earlier
square-gradient theory \cite{GT} and with corresponding results for
simple atomic fluids. We conclude in Sec.~\ref{Secsumm} with a summary
and discussion of our results.

\section{Density-functional theory for ionic fluids} \label{SecDFT}

We study the simplest model for an ionic fluid, the so-called
restricted primitive model (RPM), which consists of charged hard spheres
with equal diameters $a$ for both species. It can be considered as a
model of a molten salt, but also of an electrolyte solution with the
solvent treated as a dielectric continuum. The interaction potential
is given by 
\begin{equation}
  w_{ij}(r)=w^{HS}(r)+w_{ij}^{C}(r)
\end{equation}
where
\begin{equation}
  w^{HS}(r)=\left\{
    \begin{array}{ll}
      \infty,& r<a\\
      0, & r>a
    \end{array}
    \right.
\end{equation}
and
\begin{equation}
  w^{C}_{ij}(r)=\frac{e_i e_j}{\epsilon r} \Theta(r-a)
\end{equation}
where $r$ is the interparticle distance, $\Theta$ is the Heaviside
step function, $i,j\in\{+,-\}$, $e_+=-e_-=e$
is the charge of the particles, and $\epsilon$ is the dielectric
constant of the solvent.

The grand-canonical density functional of an
inhomogeneous fluid with number
densities $\rho_i({\bf r})$ can  be written as
\begin{eqnarray}
  \label{DFTexact}
  \Omega[\{\rho_i({\bf r})\}] & = & {\cal F}_{HS}[\{\rho_i({\bf r})\}]+\frac{1}{2}
  \sum_{i,j} \int d^3r d^3r' \rho_i({\bf r}) \rho_j({\bf r}') \int_0^1 d\alpha\,
  w_{ij}^{C}({r_{12}}) g_{ij}({\bf r},{\bf r}',\{\rho_i({\bf r})\},\alpha) \nonumber \\
  & & -\sum_i \int d^3r \mu_i \rho_i({\bf r}).
\end{eqnarray}
Here ${\cal F}_{HS}$ is the free energy functional of the corresponding
 hard sphere reference fluid,
${\bf r}_{12}={\bf r}-{\bf r}'$, $\mu_i$ is the chemical potential of chemical
species $i$, and $g_{ij}$
denotes the pair distribution function of an inhomogeneous system with
the density profiles $\rho_i({\bf r})$ and the interaction potential
$w^{HS}(r)+\alpha\, w_{ij}^{C}(r)$. The integration over $\alpha\in[0,1]$
corresponds to a straight line in potential space leading from the hard sphere
reference potential $w^{HS}$ to the full potential $w_{ij}$. For the
present model this path corresponds to a 
charging process of the ionic liquid. Although Eq.~(\ref{DFTexact}) is
formally exact \cite{Evans79}, the pair distribution function
is not known for the inhomogeneous system. We 
approximate $g_{ij}$ by the corresponding function
 $g_{ij}({\bf r}-{\bf r}',\bar\rho,\alpha)$ of a homogeneous
(bulk) liquid evaluated at an appropriate density
$\bar\rho$, for which we choose the mean value of the total densities
at the points ${\bf r}$ and ${\bf r}'$:
\begin{equation} \label{rhobar}
  \bar\rho=\bar\rho({\bf r},{\bf r}')={1\over 2} \sum_i (\rho_i({\bf r})+\rho_i({\bf r}')).
\end{equation}
A similar averaging was used in Ref.~\cite{Sokolowski} in a DFT for a
Lennard-Jones fluid in contact with a hard wall. However, in that work the
local densities were additionally  averaged over spherical regions
around ${\bf r}$ and ${\bf r}'$, which is not necessary in the present case
because there are no pronounced density oscillations at the liquid-vapor interface.
Note that the bulk pair distribution function $g_{ij}$ depends only on the
total density $\rho=\rho_++\rho_-$, which follows from the requirement
of electroneutrality, $\rho_+=\rho_-$,
in the bulk. (Strictly speaking $g_{ij}({\bf r}-{\bf r}',\bar\rho,\alpha)$
is not defined for  densities which lie within the two-phase
coexistence region of the bulk phase diagram,
but this problem does not arise for the approximate pair distribution
functions we shall actually use in our calculations below.)

By writing $g_{ij}=1+h_{ij}$ the pure Coulombic contribution to the
functional
\begin{equation} \label{FCoulomb}
  {\cal F}^{C}[\{\rho_i({\bf r})\}]={1\over 2} \sum_{i,j} \int d^3r d^3r' \rho_i({\bf r})
  \rho_j({\bf r}') w_{ij}^{C}({r_{12}})
\end{equation}
can be separated off. In the bulk, where
$\rho_+=\rho_-$, ${\cal F}^{C}$ vanishes. Similarly, due to the
symmetry of the RPM under charge inversion, the density distributions  which
minimize Eq.~(\ref{DFTexact})
at the free liquid-vapor interface should exhibit local charge
neutrality, i.e.,
$\rho_+({\bf r})=\rho_-({\bf r})={1\over 2} \rho({\bf r})$. Moreover, the symmetry of
the RPM implies that the total pair correlation functions $h_{ij}$
must satisfy $h_{++}=h_{--}$ and $h_{+-}=h_{-+}$, so that
Eq.~(\ref{DFTexact}) reduces to a functional of the total density $\rho({\bf r})$:
\begin{eqnarray}
  \label{DFTneutral}
  \Omega[\{\rho({\bf r})\}] & = & {\cal F}_{HS}[\{\rho({\bf r})\}]+\frac{1}{2} \int
  d^3r d^3r' \rho({\bf r}) \rho({\bf r}') \int_0^1 d\alpha \frac{e^2}{\epsilon
  {r_{12}}} h_D({r_{12}},\bar\rho({\bf r},{\bf r}'),\alpha) \nonumber\\
 & & -\mu \int d^3r \rho({\bf r})
\end{eqnarray}
with the difference function $h_D=(h_{++}-h_{+-})/2$ and
$\mu=(\mu_++\mu_-)/2$.  The hard sphere contribution is treated in a
local density approximation:
\begin{equation}
  \label{FHS}
  {\cal F}_{HS}[\{\rho({\bf r})\}]=\frac{1}{\beta} \int d^3r \{ \rho({\bf r})[\ln(
  \rho({\bf r})\lambda^3)-1] + \beta f_{CS}(\rho({\bf r}))\}.
\end{equation}
$\lambda$ is the thermal de Broglie wavelength and $f_{CS}$ is the
non-ideal gas part of the
free energy density of the bulk hard sphere fluid given by the accurate
Carnahan-Starling approximation \cite{CS}:
\begin{equation}
  f_{CS}(\rho)=\frac{\rho}{\beta}\frac{4\eta-3\eta^2}{(1-\eta)^2}
\end{equation}
with the packing fraction $\eta=\frac{\pi}{6}\rho a^3$. Clearly the
approximation given by Eq.~(\ref{FHS}) does not take into account the short-ranged
correlations associated with the packing constraints which give rise
to the oscillatory density profiles encountered for fluids at
walls. Nor will it account for the weak oscillations that are
predicted for low temperatures on the liquid side of the liquid-vapor
density profile in both simple \cite{EvansBulkProfile} and ionic
\cite{Raul} fluids. A non-local treatment of ${\cal F}_{HS}$ would be
required to describe such oscillatory behavior.

We employ the bulk correlation function $h_D$ given by the analytically
solvable mean spherical approximation (MSA) \cite{Waisman1,Waisman2},
which provides a reasonable description of the bulk structure and
thermodynamics of the RPM. Henderson and Smith \cite{HenSmith} have
derived an explicit expression for $h_D$:
\begin{equation}
  \label{hdMSA}
  h_D(r,\rho,T)=-\frac{\beta e^2}{\epsilon r} \frac{1}{(1+a\Gamma)^2}
  \sum_{n=1}^{[r/a]} e^{-(r-na)\Gamma} \frac{((r-na)\Gamma)^n}{(n-1)!}
  [j_{n-2}((r-na)\Gamma)-j_{n-1}((r-na)\Gamma)]
\end{equation}
where $[r/a]$ is the largest integer smaller than $r/a$, $j_n$ denotes
the spherical Bessel function of order $n$ and the inverse length
$\Gamma$ is given by $\Gamma=(\sqrt{1+2\kappa a}-1)/(2 a)$ with the
inverse Debye screening length $\kappa=(4\pi \beta e^2
\rho/\epsilon)^{1/2}$. The above expression is inconvenient for the
numerical calculation of $h_D$ at large $r$ because in this case many terms have
to be evaluated and rather large errors  may occur due to partial cancellations of
the terms. On the other hand, a very good approximation for
$h_D$ for
large $r$ can be obtained from the pole analysis of the Fourier
transform $h_D(k)$ as shown by Leote de Carvalho and Evans
\cite{Raul}. It provides the asymptotes of the form
$h_D(r\to\infty,\rho,T)=\frac{A(\rho,T)}{r} \exp(-\alpha_0(\kappa) r)$ for
$\kappa<\kappa_c=1.228/a$ and $h_D(r,\rho,T)=\frac{B(\rho,T)}{r}
\exp(-\alpha_0(\kappa) r) \cos(\alpha_1(\kappa) r+\theta(\kappa))$ for
$\kappa>\kappa_c$ with known functions $A$, $B$, $\alpha_0$,
$\alpha_1$, and $\theta$.  Using these methods we have derived similar, but slightly more complicated
expressions for $\frac{\partial}{\partial\rho} h_D$
and $\frac{\partial^2}{\partial \rho^2} h_D$, which are needed to compute
the phase diagram and the correlation length (see Sec.~\ref{Secbulk}).
These asymptotic approximations have been used in the numerical
calculations for $r/a>12$.

\section{Bulk phase diagram and correlation length} \label{Secbulk}

For a constant density $\rho({\bf r})=\rho$ the density functional 
yields a Helmholtz free energy density
\begin{equation}
  \label{Fbulk}
  \frac{F(\rho)}{V}=\frac{\rho}{\beta}(\ln(\rho\lambda^3)-1) 
  + f_{CS}(\rho)+{1\over 2} \rho^2 w_0(\rho)
\end{equation}
with
\begin{equation}
  \label{w0}
  w_0(\rho)=\frac{e^2}{\epsilon} \int d^3 r\, r^{-1} \int_0^1 d\alpha\,
  h_D(r,\rho,\alpha). 
\end{equation}
Since $h_D$ depends only on the dimensionless quantities
$\frac{\beta e^2}{\epsilon a}$ and $r/a$, the charging integration
with $e^2(\alpha)=\alpha e^2$ can be replaced by an integration over
$\beta$:
\begin{equation}
  \label{w0def}
  w_0(\rho)=\frac{e^2}{\beta\epsilon} \int d^3r\, r^{-1} \int_0^\beta d\beta'
  h_D(r,\rho,\beta').
\end{equation}
This result
 demonstrates that the free energy given by Eqs.~(\ref{Fbulk}) and (\ref{w0}) is
equal to that obtained by the so-called energy route starting from the
MSA pair correlation function. Thus the bulk phase diagram is identical to
the usual MSA phase diagram as discussed, e.g., in
Refs.~\cite{Guillot,LevinPhysica,GT,Raul,Raul2}. The explicit expression for $w_0(\rho)$ is
\cite{Waisman2} 
\begin{equation} \label{w0MSA}
  {1\over 2} \rho^2 w_0(\rho)=-\frac{1}{4\pi\beta a^3}((\kappa a)^2+2\kappa
  a+\frac{2}{3}-\frac{2}{3} (1+2\kappa a)^{3/2}).
\end{equation}
For the reasons discussed in Sec.~\ref{Secinterf} below, we
also consider the functional which arises when we refrain from the integration over $\alpha$ and set $\alpha=1$
instead, which amounts to replacing the excess free energy
${\cal F}_{ex}={\cal F}-{\cal F}_{HS}$ due to the Coulomb interactions by the corresponding
internal energy $U_{ex}=\frac{\partial}{\partial\beta} (\beta
{\cal F}_{ex})$ (see Eq.~(\ref{w0def})). We call this approximation scheme
MSA1. In this case Eq.~(\ref{w0MSA}) is replaced by
\begin{equation}
  \label{w0MSA1}
  {1\over 2} \rho^2 w_0(\rho)=-\frac{1}{4\pi\beta a^3}(\kappa a+(\kappa
  a)^2-\kappa a \sqrt{1+2 \kappa a}).
\end{equation}
Note that, in contrast to Eq.~(\ref{w0MSA}), this expression does not reduce to
the exact Debye-H\"uckel result ${1\over 2} \rho^2
w_0(\rho)=-\kappa^3/(12\pi \beta )$ in the low-density ($\kappa\to 0
$) limit. However, it differs only by a factor of $\frac{3}{2}$.

The liquid-vapor phase coexistence curves, which follow from the free energy by
the usual double tangent construction, are shown in Fig.~\ref{fig:phas}
for the approximations MSA and MSA1, together with Gibbs ensemble Monte Carlo
simulation results \cite{Pana,Smittripel}. 
The results are given in terms of the reduced density $\rho^\ast=\rho
a^3$ and the reduced temperature $T^\ast=k_B T \epsilon a/e^2$. The
phase diagram of the RPM is characterized by small values of the reduced
critical density and temperature (for comparison the Lennard-Jones
fluid  has $T^\ast_c\simeq 1.3$ and $\rho^\ast_c\simeq 0.3$ in Lennard-Jones
reduced units) and by a strong
asymmetry of the coexistence curve. At low temperatures the vapor phase
becomes extremely dilute, e.g., at $T^\ast=0.04\simeq T^\ast_c/2$ one has 
$\rho^\ast_v\simeq 10^{-7}$ within the  MSA. It is well known that the MSA
overestimates the critical temperature and grossly underestimates the
critical density. The cruder approximation MSA1 predicts an even higher
critical temperature (MSA: $T^\ast_c=0.0786$, MSA1: $T^\ast_c=0.0846$) and
smaller critical density (MSA: $\rho^\ast_c=0.0145$, MSA1:
$\rho^\ast_c=0.0086$). Simulation results for the coexistence curve have
undergone substantial revision during recent years (see,
e.g., Refs.~\cite{FisherJSP,FisherNorwich}); the latest estimate of the
critical point by Caillol et al. \cite{CaillolRPMMC}, using a
mixed-field finite size analysis of Monte Carlo data, gives
$T^\ast_c=0.0488\pm 0.0002$ and $\rho^\ast_c=0.080\pm0.005$. Clearly
quantitatively reliable results cannot be expected from either MSA or
MSA1. Both fail to take proper account of the effects of ion pairing
which is known to be very strong in the RPM at low densities and which
is believed to influence strongly the location of the critical point
\cite{Guillot,FisherJSP,FisherNorwich}. Since the differences between
the coexistence curves obtained from MSA and MSA1 are relatively minor
we do not expect to introduce significant additional errors by
using MSA1.
We note that
the main contributions to the integration over $\alpha$ in Eq.~(\ref{w0})
stem from the region near $\alpha=1$.

The chemical potential $\mu_0$ at coexistence is, for a given temperature,
\begin{equation}
  \label{mu}
  \mu_0=\frac{\partial}{\partial\rho}\left.\frac{F(\rho)}{V}\right|_{\rho_l} 
   = \frac{\partial}{\partial\rho}\left.\frac{F(\rho)}{V}\right|_{\rho_v} 
\end{equation}
where $\rho_l$ and $\rho_v$ are the densities of liquid and vapor,
respectively. 
Knowledge of $\mu_0(T)$ is a prerequisite for the subsequent
analysis of the liquid-vapor interface.

We now turn attention to the bulk correlation
length that follows from the present theory. As we shall see in the
next section, the width of the interface is governed by the
correlation length of the bulk liquid.
Every density functional
ansatz  defines a direct correlation function via
twofold functional differentiation of the intrinsic Helmholtz free
energy functional ${\cal F}[\{\rho_i({\bf r})\}]$ with respect to density
\cite{Evans79}:
\begin{equation}
  \label{callg}
  c_{ij}({\bf r},{\bf r}',\{\rho_i({\bf r})\})=-\beta\frac{\delta^2
  {\cal F}[\{\rho_i({\bf r})\}]} {\delta\rho_i({\bf r}) \delta\rho_j({\bf r}')} +
  \frac{\delta_{ij} \delta({\bf r}-{\bf r}')}{\rho_i({\bf r})}.
\end{equation}
From the bulk limit of  this function a pair distribution function $h_{ij}$
can then be obtained via the Ornstein-Zernike equation. The
results will be different from the $h_{ij}$ used for the construction
of the functional, in our case those of the MSA. This is an important
advantage of the density-functional approach. We require the
correlation function for fluctuations of the total density
$h_\rho^{(b)}=(h_{++}+h_{+-})/2$ to exhibit the conventional
Ornstein-Zernike behavior, i.e., $h_\rho^{(b)}$ should become
long-ranged near the critical point, with an exponential decay
described by a diverging correlation length. Such a behavior reflects
the net inter-particle interaction which is necessary to obtain
liquid-vapor coexistence. However, as remarked in the Introduction,
the original MSA result for $h_\rho^{(b)}$ is simply that of
hard-spheres (in the PY approximation), so there is no manifestation
of attraction in this particular combination of the correlation
functions. We define the second-moment correlation length
$\xi$ for fluctuations of the total density by the expansion of the
structure factor:
\begin{equation}
  \label{xidef}
  S(k,\rho)=1+\rho h_\rho^{(b)}(k,\rho)=S(0,\rho)/(1+\xi^2 k^2+\cdots),
\end{equation}
where $k$ is the wave number. This correlation length
can be obtained rather easily
if one again assumes $\rho_+({\bf r})=\rho_-({\bf r})={1\over 2} \rho({\bf r})$ and
considers ${\cal F}=\Omega+\mu\int d^3r \rho({\bf r})$ (see Eq.~(\ref{DFTneutral})) as a functional of the total
density $\rho({\bf r})$. Using the bulk limit $c_\rho^{(b)}$ of the function
\begin{equation}
  c_\rho({\bf r},{\bf r}',\{\rho({\bf r})\})=-\beta\frac{\delta^2
  {\cal F}[\{\rho({\bf r})\}]} {\delta\rho({\bf r}) \delta\rho({\bf r}')} +
  \frac{\delta({\bf r}-{\bf r}')}{\rho({\bf r})}
\end{equation}
$\xi$ is given by
\begin{equation}
  \label{xi}
  \xi^2=\frac{1}{6} \frac{\int d^3r\, r^2 c_\rho^{(b)}(r,\rho)}
  {1/\rho-\int d^3r c_\rho^{(b)}(r,\rho)},
\end{equation}
which follows  straightforwardly from the definition given by Eq.~(\ref{xidef}) and the
Ornstein-Zernike equation
$h_\rho^{(b)}(k,\rho)=c_\rho^{(b)}(k,\rho)/(1-\rho
c_\rho^{(b)}(k,\rho))$.
Within the MSA1 approximation one finds
\begin{eqnarray}
  \label{cbulk}
  \beta^{-1} c_\rho^{(b)}(r,\rho) & = &
  -\delta({\bf r})\frac{\partial^2}{\partial\rho^2} f_{CS}(\rho)
   -\frac{e^2}{\epsilon r} [ h_D(r,\rho)+\rho
  \frac{\partial}{\partial\rho} h_D(r,\rho)+\frac{1}{4} \rho^2
  \frac{\partial^2}{\partial\rho^2} h_D(r,\rho)] \nonumber \\
  & & {}-\delta({\bf r}) \int d^3r' \frac{e^2}{\epsilon r'} [\rho
  \frac{\partial}{\partial\rho} h_D(r',\rho)+\frac{1}{4} \rho^2
  \frac{\partial^2}{\partial\rho^2} h_D(r',\rho)].
\end{eqnarray}
Note that the equivalent result for a simple atomic fluid was derived
in Ref.~\cite{Schirmacher}.
The resulting correlation lengths of the coexisting liquid and vapor
phase are plotted in Fig.~\ref{fig:korr}. In accordance with the
mean-field character of the present approach they diverge near the critical
point proportional to $(T_c-T)^{1/2}$. $\xi$ also diverges upon
approaching the spinodals, which are determined by $\frac{\partial^2
  F(\rho)}{\partial \rho^2}=0$, or, equivalently, by the vanishing of
the denominator in Eq.~(\ref{xi}). Upon lowering the temperature the
correlation length in the liquid phase decreases  to about one
particle diameter, whereas that of the vapor phase increases again as a
consequence of the reduced screening at very low densities. 

It is instructive to examine the limiting behavior as $\rho\to 0$. The exact
low density behavior of the second moment correlation length is
believed to be \cite{LeeLetter}
\begin{equation} \label{limr0ex}
  \xi=\frac{1}{4} \left(\frac{\beta e^2}{36 \pi \epsilon
  \rho}\right)^{1/4} [1+O(\rho^{1/2})],
\end{equation}
a result which also follows from the HNC approximation \cite{Ennis,LeeLetter}.
If one assumes that for low densities the MSA correlation function $h_D$
tends to $-\frac{\beta e^2}{\epsilon r} e^{-\kappa r}$ all the algebra
can be performed analytically and the MSA1 yields in leading order
\begin{equation} \label{limr0MSA1}
  \xi=\left(\frac{7}{192\pi}\frac{\kappa}{\rho}\right)^{1/2}=\frac{\sqrt{7}}{4} \left(\frac{\beta e^2}{36 \pi \epsilon \rho}\right)^{1/4},
\end{equation}
which is in accordance with our numerical results. Thus, for $\rho\to 0$
the MSA1 correlation length diverges with the correct power law but the
amplitude is too large by a factor of $\sqrt{7}$. We can also carry
through the algebra using the original functional Eq.~(\ref{DFTneutral})
(including the $\alpha$ integration). In this case we obtain
\begin{equation} \label{limr0MSA}
  \xi=\frac{\sqrt{14}}{4} \left(\frac{\beta e^2}{36 \pi \epsilon \rho}\right)^{1/4}
\end{equation}
for $\rho\to 0$. Once again the power law is correct but the amplitude
is again too large. These results suggest that the functionals capture the
essential features of density-density correlations in the $\rho\to 0$
limit. This is a non-trivial observation. Recall that GMSA, which
assumes a particular (single Yukawa) form for the non-Coulomb part of
the direct correlation functions $c_{ij}(r)$ and then enforces
consistency among three routes to bulk thermodynamic functions, yields
a finite value as $\rho\to 0$
\cite{LeeLetter}.  Thus our present MSA1 and MSA
results, Eqs.~(\ref{limr0MSA1}) and (\ref{limr0MSA}), improve upon the
GMSA in the low-density limit. Finally we should emphasize that the
results described here refer to the second moment correlation length
$\xi$ as defined by Eqs.~(\ref{xidef}) or (\ref{xi}). The so-called
true correlation length $\xi_\infty$ that determines the ultimate
exponential decay of the density-density correlation function
$h_\rho^{(b)}(r,\rho)$ is determined by the poles of the Fourier
transform $h_\rho^{(b)}(k,\rho)$ \cite{Raul}. In general $\xi_\infty$
differs from $\xi$; they become identical (within mean-field theory)
in the vicinity of the critical point. Calculating the poles from the
present DFT approach is not straightforward as the Fourier transform
of Eq.~(\ref{cbulk}) cannot be performed analytically, so we cannot easily
compute $\xi_\infty$. However, we are able to compare our results for
$\xi$ with $\xi_{\infty,GMSA}$ obtained in Ref.~\cite{Raul}. For subcritical
liquid-like states which turn out to be relevant for the interface,
the  correlation lengths $\xi_{MSA}$ and $\xi_{MSA1}$ both differ by
less than a factor of
two from $\xi_{\infty,GMSA}$. In the very low density vapor phase the exact limiting behavior
is $\xi_\infty\simeq 1/(2\kappa)$ \cite{Ennis,LeeLetter} which diverges
as $(T/\rho)^{1/2}$ for $\rho\to 0$, i.e., faster than the second
moment correlation length given by Eq.~(\ref{limr0ex}).

In summary we conclude that the present density-functional theory
provides a reasonable description of the long-wavelength (small $k$)
behavior of
density-density correlations in the bulk RPM. This is a necessary
prerequisite for a reliable treatment of the liquid-vapor interface.

\section{Liquid-vapor interface} \label{Secinterf}

\subsection{Density profiles}

The average density profile of the liquid-vapor interface varies only in
the direction normal to the interface which we take as the $z$
direction. In the following we assume that the bulk liquid density
$\rho_l$ is
obtained for $z\to-\infty$ and the bulk vapor density $\rho_v$ for
$z\to\infty$. The application of the density-functional theory
outlined in Sec.~\ref{SecDFT} to this situation allows one to
carry out the lateral integrations in Eq.~(\ref{DFTneutral}) yielding the
following  grand-canonical functional per surface area $A$ for the
total density $\rho(z)=2\rho_-(z)=2\rho_+(z)$:
\begin{equation}
  \label{DFTlg}
  \frac{1}{A}\Omega[\{\rho(z)\}]=\frac{1}{A} {\cal F}_{HS}[\{\rho(z)\}]+{1\over 2}
  \int dz dz'\,\rho(z) \rho(z') w(z-z',\bar\rho(z,z')) - \mu_0 \int dz\,\rho(z)
\end{equation}
with
\begin{equation}
  \label{wdef}
  w(z_{12},\rho)=\frac{2\pi e^2}{\epsilon} \int_{|z_{12}|}^\infty dr
  \int_0^1 d\alpha\,h_D(r,\rho,\alpha).
\end{equation}

For densities  not too large the function $h_D$ exhibits its slowest
decay as function of $r$ (lowest value of $\alpha_0(\kappa)$, the
imaginary part of the leading pole) for small values of the charge, i.e., for small $\alpha$ (see
Fig.~4(b) in Ref.~\cite{Raul}). Therefore, in contrast to the bulk free
energy, the large distance behavior of $w(z_{12},\rho)$ is dominated
by the contributions from small $\alpha$. In this region the
correlation function is expected to take on the Debye-H\"uckel form
\begin{equation}
  \label{hdlim}
  h_D(r,\rho,\alpha)\simeq -\frac{\beta e^2}{\epsilon r} \alpha
  e^{-\kappa \sqrt{\alpha} r}
\end{equation}
which also follows from the small coupling expansion of the decay length and of the
expression for the
amplitude obtained from the pole analysis \cite{Raul} mentioned in
Sec.~\ref{Secbulk}. From Eqs.~(\ref{wdef}) and (\ref{hdlim}) one finds
\begin{equation}
  \frac{\partial}{\partial z} w(z\to\infty,\rho) \simeq 
  \frac{2\pi \beta e^4}{\epsilon^2 z} \int_0^1 d\alpha\, \alpha
  e^{-\kappa\sqrt{\alpha} z}
\end{equation}
and hence
\begin{equation}
  w(z\to\infty,\rho) \simeq -\frac{6\pi \beta e^4}{\epsilon^2 \kappa^4
  z^4}.
\end{equation}
Thus  $w$, which corresponds to the effective interaction between
two fluid layers a distance $z$ apart, is predicted to decay
algebraically with the same exponent as for a
Lennard-Jones fluid, where the interatomic potential decays as
$r^{-6}$. This unexpected result implies an algebraic decay for the
density profiles, too. More specifically one would expect
$d\rho/dz\sim-|z|^{-4}$ as $z\to\pm\infty$. This behavior can be
traced back to the fact that within the MSA density-functional theory
the bulk direct correlation function $c_\rho^{(b)}(r;\rho)$, which is given by Eq.~(\ref{cbulk}) with
$h_D(r,\rho)$ replaced by $\int_0^1 d\alpha\, h_D(r,\rho,\alpha)$,
also decays algebraically. Using an  argument similar to the one above one
finds $c_\rho^{(b)}(r,\rho)\sim r^{-6}$ as $r\to\infty$, which is the
same as for a Lennard-Jones fluid. This would, in turn, imply that
$h_\rho^{(b)}(r;\rho)$ as obtained from $c_\rho^{(b)}(r,\rho)$ via the
Ornstein-Zernike equation should also decay as $r^{-6}$. In other
words the theory generates effective potentials for density-density
correlations which mimic those of genuine dispersion forces. Whilst
this observation is certainly intriguing it is likely to be an
artefact of the present MSA scheme. The bulk pair correlation function
$h_\rho^{(b)}(r;\rho)$ should decay exponentially for the RPM
\cite{Ennis,Raul}. Moreover, general arguments
for the decay of density profiles near walls
\cite{EvansBulkProfile,Raul,Attard93} show that the ultimate decay of $\rho(z)$
should mimic that of $h_\rho^{(b)}$. For example $\rho_l-\rho(z)$ should decay as
$\exp(z/\xi_\infty)$ as $z\to-\infty$, where $\xi_\infty$ is the true
correlation length of the bulk liquid. If, on the other hand, we
ignore the $\alpha$ integration and set 
$\alpha=1$  (see also the discussion in
Sec.~\ref{Secbulk}) then $-\frac{\partial w(z,\rho)}{\partial z}\sim
h_D(z,\rho)$ and there is no longer an algebraic decay of the direct
correlation function and of the
profile. Henceforward we describe results based on both MSA and MSA1.

By functional differentiation of Eq.~(\ref{DFTlg}) one derives the
Euler-Lagrange equation
\begin{equation}
  \label{ELG}
  \log \rho(z)\lambda^3+f'_{CS}(\rho(z))=\mu_0-p(z)
\end{equation}
with
\begin{equation}
  \label{pz}
  p(z)=\int dz' \rho(z')\left[w(z-z',\bar\rho(z,z'))+{1\over 2} \rho(z)
  w'(z-z',\bar\rho(z,z'))\right].
\end{equation}
(The primes on $f_{CS}$ and $w$ denote derivatives with respect to the
density.) In contrast to similar theories for Lennard-Jones \cite{EvansLesHouches}
or dipolar \cite{Frodl} fluids the kernel in Eq.~(\ref{pz}) depends on $z$
and $z'$ separately due to the density dependence of the
approximated pair distribution function. However,  in the present
case one can still show that
if a profile $\rho(z)$ solves Eq.~(\ref{ELG}) any shifted profile $\rho(z-z_0)$
will be a solution as well, because the position of the interface is not
fixed by
the boundary conditions. In order to select a unique profile we have
imposed the additional requirement
$\rho(z=0)=(\rho_l-\rho_v)/2$ and have shifted the profiles
accordingly during the numerical solution.

 Equation (\ref{ELG}) was solved
using the usual Picard iteration scheme. In each iteration $(n)$ the
function $p(z)$ is calculated from the density profile $\rho_{(n)}(z)$
in the region $z\in [-L/2,L/2]$
and a new $\rho_{new}(z)$ is obtained by numerical inversion of
Eq.~(\ref{ELG}) for each value of $z$. In order to achieve convergence the
profile for the next iteration is calculated according to the mixing
rule
\begin{equation}
  \rho_{(n+1)}=\omega \rho_{new}+(1-\omega) \rho_{(n)}
\end{equation}
where typically $\omega=0.2$. The process is repeated until $\max_z|\rho_{(n+1)}(z)-\rho_{(n)}(z)|$ is smaller than a
prescribed accuracy. Since the functions $w(z_{12},\rho)$ and
$w'(z_{12},\rho)$ have discontinuous derivatives at $z_{12}=\pm a$ the integration in
Eq.~(\ref{pz}) is divided into the three subintervals $[-L/2,z-a]$,
$[z-a,z+a]$, and $[z+a,L/2]$. The Milne rule, which effectively
interpolates the integrand piecewise by third order polynomials, is
used for all integrations. Asymptotic contributions to $p(z)$ from the regions
$|z'|>L/2$ are calculated by replacing $\rho(z')$ by the bulk limits
$\rho(z'<-L/2)=\rho_l$ and $\rho(z'>L/2)=\rho_v$ and integrating
numerically with $w(|z-z'|>20 a)=0$.

The functions $w$ and $w'$ are given by integrals over the functions $h_D$ and
$h_D'$ whose evaluation already requires significant numerical efforts. Since
$w(z_{12},\rho)$ and $w'(z_{12},\rho)$ have to be evaluated many times during
the iteration scheme the algorithm can be accelerated considerably by using
two-dimensional spline interpolations for these functions, which require their
evaluation only at, e.g., $100\times100$ grid points before the main algorithm
starts.

In Fig.~\ref{fig:profil} we plot the density profiles obtained from MSA1
for a series of different
temperatures. Upon increasing the temperature towards the critical point
($T^\ast_c=0.08465$) the interface broadens and the density difference
$\rho_l-\rho_v$ between
the phases decreases. These obvious features can be incorporated by
introducing a scaling function $\rho_{scl}(z/\xi,T)$ with
$\rho_{scl}(\pm\infty,T)=\mp 1$:
\begin{equation}
  \label{rhoscl}
  \rho(z,T)={1\over 2} (\rho_l+\rho_v)+{1\over 2} (\rho_l-\rho_v) \rho_{scl}(z/\xi,T).
\end{equation}
For $T\to T_c$ the scaling function should reduce to a universal
function of the single variable $z/\xi$,
which --- within mean-field theory --- is given by
\begin{equation}
  \label{tanh}
  \rho_{scl}(z/\xi,T\to T_c)=-\tanh\frac{z}{2\xi}.
\end{equation}
Such a plot is given in Fig.~\ref{fig:rhoscl} using the second moment
correlation length $\xi_l(T)$ for the bulk
liquid phase, obtained from Eqs.~(\ref{xi}) and (\ref{cbulk}). This demonstrates that the scaling behavior
remains valid even relatively far outside the critical region.
In order to make  comparison with the behavior of  simple atomic fluids
we present  in Fig.~\ref{fig:rholj} the density profiles of a Lennard-Jones fluid scaled in the
same manner. These results have been obtained using
the density-functional ansatz of Frodl and Dietrich \cite{Frodl}. 
At low temperatures the profiles of the ionic fluid deviate weakly
from antisymmetry, approaching the bulk
limit faster on the vapor side than on the liquid side. The same trend can be observed in the
Lennard-Jones system, where it is even more pronounced: compare profiles
corresponding to the same reduced temperature $t=1-T/T_c$. In the case
of a Lennard-Jones fluid the deviation from antisymmetry is attributed \cite{Lu} to
the fact that $\xi_v<\xi_l$ at low temperatures. The situation is
different for the ionic case where, away from $T_c$, $\xi_v\gg\xi_l$
(see Fig.~\ref{fig:korr}). This would imply that within a square
gradient theory there should be a {\it slower} decay of the profiles on the
{\it vapor} side. The results of Ref.~\cite{GT} are in accordance with this
expectation. However, compared with the square gradient theory our present
non-local theory clearly predicts the opposite asymmetry: the decay on
the vapor side is {\it faster} than on the liquid side, although
$\xi_v\gg\xi_l$. In order to understand these features it is important to distinguish between the
intermediate range decay that is apparent from
Figs.~\ref{fig:profil}--\ref{fig:rholj} and the ultimate asymptotic
decay into the bulk. As mentioned previously, for the ionic fluid the
ultimate decay should be exponential with a decay length equal to the
true correlation length $\xi_\infty$ of the bulk. We attempted to
analyze the numerical results for the tails of our profiles but this
is  rather difficult
 because the variations are very small. When curves of the form $C
\exp(-|z|/\xi)$ or $(C/z) \exp(-|z|/\xi)$ are fitted to the tails of
the MSA1 profiles
one finds larger values for $\xi$ on the vapor side, as expected, but
the actual values 
depend very strongly on the $z$ interval used for the fitting. Finally
we note that the density profiles in a Lennard-Jones fluid approach,
as a function of temperature, the universal scaling function
monotonically from below on the vapor side and from above on the
liquid side (see Fig.~\ref{fig:rholj}). On the other hand in the ionic
fluid the density profile on the vapor side is below the universal
profile at low temperature but is above it for large $z$ close and $T$
to $T_c$ (see
Fig.~\ref{fig:rhoscl} for $t=1-T/T_c=0.0313$). We interpret this as a consequence
of the facts that $\xi_v>\xi_l$ and that the vapor becomes more important
near $T_c$ due to its increasing density. At the same reduced
temperature the scaled interfacial profiles in a Lennard-Jones fluid
are steeper than those of the ionic fluid.

We define the interfacial width $\delta$  as
\begin{equation}
  \label{width}
  \delta=-(\rho_l-\rho_v)/\left.\frac{d\rho}{dz}\right|_{z=0}.
\end{equation}
For the  tanh profile in Eq.~(\ref{tanh}) one has
$\delta=4\xi$, so that in the critical region $\delta/4$ should
diverge in the same way as the correlation length. Remarkably,
Fig.~\ref{fig:korr} shows that at {\it all} temperatures this width
is determined by the {\it liquid} correlation length $\xi_l$. At
$T^\ast=0.03$, which is a little higher than the estimated triple point
temperature \cite{Smittripel}, $\delta/a\simeq3.0$. This value is
slightly higher than the estimates of the equivalent ratio for simple
atomic fluids near their triple points \cite{RW}.  Also shown in
Fig.~\ref{fig:korr} are the results of Ref.~\cite{GT} which
exhibit a similar variation with $T$  but at low $T$ the widths are
about a factor of two smaller. The '10--90'
interfacial thickness used
in Ref.~\cite{GT} is defined as the distance over which the total density
changes from $\rho_v+0.9(\rho_l-\rho_v)$ to
$\rho_v+0.1(\rho_l-\rho_v)$. For the tanh profile it gives a width
that is larger by a factor ${\rm artanh}(0.8)=1.0986$ than the
quantity $\delta$
defined in Eq.~(\ref{width}). The data taken from Ref.~\cite{GT} have been
divided by this factor. For
the actual profiles shown in Fig.~\ref{fig:profil} the two definitions
for the width yield values which differ by about 10\%, too.

The density profiles obtained from the MSA (including the $\alpha$
integration) are, for a given value of $t=1-T/T_c$, similar to
those obtained from MSA1. When plotted in terms of scaled variables,
Eq.~(\ref{rhoscl}), the profiles from the two theories can hardly be
distinguished on the scale of Fig.~\ref{fig:rhoscl}. Moreover,
identifying the ultimate algebraic decay from the numerical results
for the tails of the profiles was not feasible. The interfacial widths
$\delta$ calculated from the MSA profiles are close to $4\xi_l$, with
$\xi_l$ obtained within the MSA approach. These widths are in turn
close to those from MSA1, provided we make comparison at the same
value of $t$.

\subsection{Surface tension}

In this subsection we  present results of calculations of the liquid-vapor
surface tension $\gamma$. First we derive an expression for $\gamma$ in terms of the density profile
$\rho(z)$ which is convenient for numerical work. Since it is
necessary to cut off
 the system at $z=\pm L/2$ this generates artificial fluid-vacuum surface
tensions $\gamma_{l,vac}$ and $\gamma_{v,vac}$ which must be subtracted in order to obtain the liquid-vapor
surface tension. Therefore $\gamma$ is given by
\begin{equation}
  \label{surfdef}
  \gamma=\lim_{L\to\infty}\left[\frac{\Omega(L)}{A}-\frac{\Omega_{bulk}}{A}
  \right] - \gamma_{l,vac}-\gamma_{v,vac}
\end{equation}
where $\Omega(L)$ is the grand-canonical functional evaluated for a
finite system of size $L$.
It is convenient to express the bulk free energy (Eq.~(\ref{Fbulk})) as a double
integral over $z$ and $z'$ analogous to Eq.~(\ref{DFTlg}). A straightforward
calculation yields
\begin{equation}
  \label{bulkint}
  \int_0^{L/2} dz \int_0^{L/2} dz'\, w(z-z',\rho)=\frac{L}{2} w_0(\rho) -
  2\int_0^{L/2} dz\, t(z,\rho)
\end{equation}
with $t(z,\rho)=\int_z^\infty dy\,w(y,\rho)$ and, from Eqs.~(\ref{w0def}) and
(\ref{wdef}),
\begin{equation}
  w_0(\rho)=2 \int_0^\infty dy\,w(y,\rho)=2 t(0,\rho).
\end{equation}
The surface tensions with the vacuum follow from assuming constant density profiles
$\rho(z)=\rho_b$, with $b=v,l$, which leads to (there are two equal interfaces)
\begin{equation}
  \gamma_{b,vac}={1\over 2} \lim_{L\to\infty}{1\over 2} \rho_b^2 \left[ 
  \int_{-L/2}^{L/2} dz \int_{-L/2}^{L/2} dz'\, w(z-z',\rho_b)-L w_0(\rho_b)
  \right].
\end{equation}
Using Eq.~(\ref{bulkint}) and the symmetry $w(-y)=w(y)$ gives
\begin{eqnarray}
  \label{surfvac}
  \gamma_{b,vac} & = & {1\over 2} \rho_b^2 \lim_{L\to\infty} \left[ 
  \int_0^{L/2} dz \int_{-L/2}^0 dz'\, w(z-z',\rho_b)- 2 \int_0^{L/2} dz\,
  t(z,\rho_b)   \right] \nonumber \\
  & = & {1\over 2} \rho_b^2 \lim_{L\to\infty} \left[ \int_0^{L/2} dz
  (t(z,\rho_b)-t(L/2+z,\rho_b)) - 2 \int_0^{L/2} dz\,
  t(z,\rho_b)   \right] \\
  & = & -{1\over 2} \rho_b^2 \int_0^\infty dz\, t(z,\rho_b). \nonumber
\end{eqnarray}
Thus the final expression for the surface tension follows by inserting
Eqs.~(\ref{Fbulk}), (\ref{bulkint}), and (\ref{surfvac}) into Eq.~(\ref{surfdef}):
\begin{eqnarray}
  \label{surfres}
  \gamma & = & \int_{-\infty}^\infty dz [f_{HS}(\rho(z))-f_{HS}(\rho_{SK}(z))]
  -\mu_0 \int_{-\infty}^\infty dz [\rho(z)-\rho_{SK}(z)] \nonumber \\
  & & {}+{1\over 2} \int_0^\infty dz \int_0^\infty dz' \left[\rho(z)\rho(z')
  w(z-z',\bar\rho(z,z'))-\rho_v^2 w(z-z',\rho_v)\right]  \\
  & & {}+{1\over 2} \int_{-\infty}^0 dz \int_{-\infty}^0 dz' \left[\rho(z)\rho(z')
  w(z-z',\bar\rho(z,z'))-\rho_l^2 w(z-z',\rho_l)\right] \nonumber \\
  & & {}+\int_{-\infty}^0 dz \int_0^\infty dz' \rho(z) \rho(z') w(z-z',\bar\rho(z,z')) -{1\over 2}
  \rho_v^2 \int_0^\infty dz\,t(z,\rho_v)-{1\over 2}
  \rho_l^2 \int_0^\infty dz\,t(z,\rho_l) \nonumber
\end{eqnarray}
with the sharp-kink profile
\begin{equation}
  \rho_{SK}(z)=\left\{
    \begin{array}{ll}
      \rho_l, & z<0 \\
      \rho_v, & z>0
    \end{array}
\right.
\end{equation}
and $f_{HS}(\rho)=\frac{\rho}{\beta}[\ln(\rho\lambda^3)-1]+f_{CS}(\rho)$. All
integrals occuring in Eq.~(\ref{surfres}) are convergent. 

Since the surface tension is the surface excess grand potential per
unit area, an alternative starting point is the formula
\begin{equation} \label{surfalter}
  \gamma=\int_{-\infty}^\infty dz (\omega(z)+p)
\end{equation}
with the grand-canonical potential density
\begin{equation}
  \omega(z)=f_{HS}(z)-\mu_0\rho(z)+{1\over 2} \rho(z) \int_{-\infty}^\infty dz'
  \rho(z') w(z-z',\bar\rho(z,z'))
\end{equation}
and the vapor pressure
\begin{equation}
  p=-f_{HS}(\rho_b)+\mu_0\rho_b-{1\over 2} \rho_b^2 w_0(\rho_b),\qquad b=l,v.
\end{equation}
It is straightforward to show that Eq.~(\ref{surfalter}) also leads to
Eq.~(\ref{surfres}).

The surface tensions obtained from the MSA1 and MSA theories are
displayed in Fig.~\ref{fig:surf} as a function of $T/T_c$ and are
compared with the results of the square-gradient theory
\cite{GT}. Since MSA1 has a higher critical temperature $T_c$ than MSA
we plot the ratio $\tilde\gamma=\gamma a^2 /(k_B T_c)$ rather than the
reduced tension $\gamma^\ast=\gamma\epsilon a^3/e^2$, which is
displayed in Ref.~\cite{GT}. The MSA results for $\tilde\gamma$ are larger than the values
predicted by MSA1 which in turn are larger than those obtained from
the square-gradient theory, with the differences becoming
larger at low temperatures. Near $T_c$ all three theories yield the
standard mean-field critical behavior, i.e., $\tilde\gamma\sim
(1-T/T_c)^{3/2}$. At lower temperatures the square-gradient theory
gives a nearly linear variation of $\tilde\gamma$ with $T/T_c$,
whereas in MSA1 the $(1-T/T_c)^{3/2}$ behavior appears to
persist further from $T_c$, which is consistent with the earlier
observation that the density profiles follow the scaling behavior even
for temperatures well removed from $T_c$.

The present MSA1 results predict
$\tilde\gamma(T/T_c=0.6)\simeq0.09$. This estimate should be
contrasted with the corresponding result for a Lennard-Jones fluid
where simulations and theories yield $\tilde\gamma(T/T_c=0.6)\simeq0.6$
\cite{RW}. The physical origin of such a large difference is not
obvious and it is important to ask whether it is an artefact of the
present approximations. Although our theories overestimate $T_c$ they
should make a compensating overestimation of the surface tension
$\gamma$. It is conceivable that since MSA1 and MSA grossly
underestimate, for a given value of $T/T_c$, the simulation results
for the liquid densities at coexistence (see Fig.~\ref{fig:phas}) they
also underestimate the magnitude of $\gamma$. On the other hand, if
one takes experimental data for the surface tension of molten alkali
halides near the melting points and makes some reasonable estimates of
the (average) diameter $a$ the values of the resulting reduced
tension $\gamma^\ast$ \cite{GT} are similar to those obtained by
extrapolating the present results to the appropriate reduced
temperatures $T^\ast$. This would suggest that the present theories yield
at least the correct order of magnitude for the surface tenstion
$\gamma$.

\section{Summary and Discussion} \label{Secsumm}

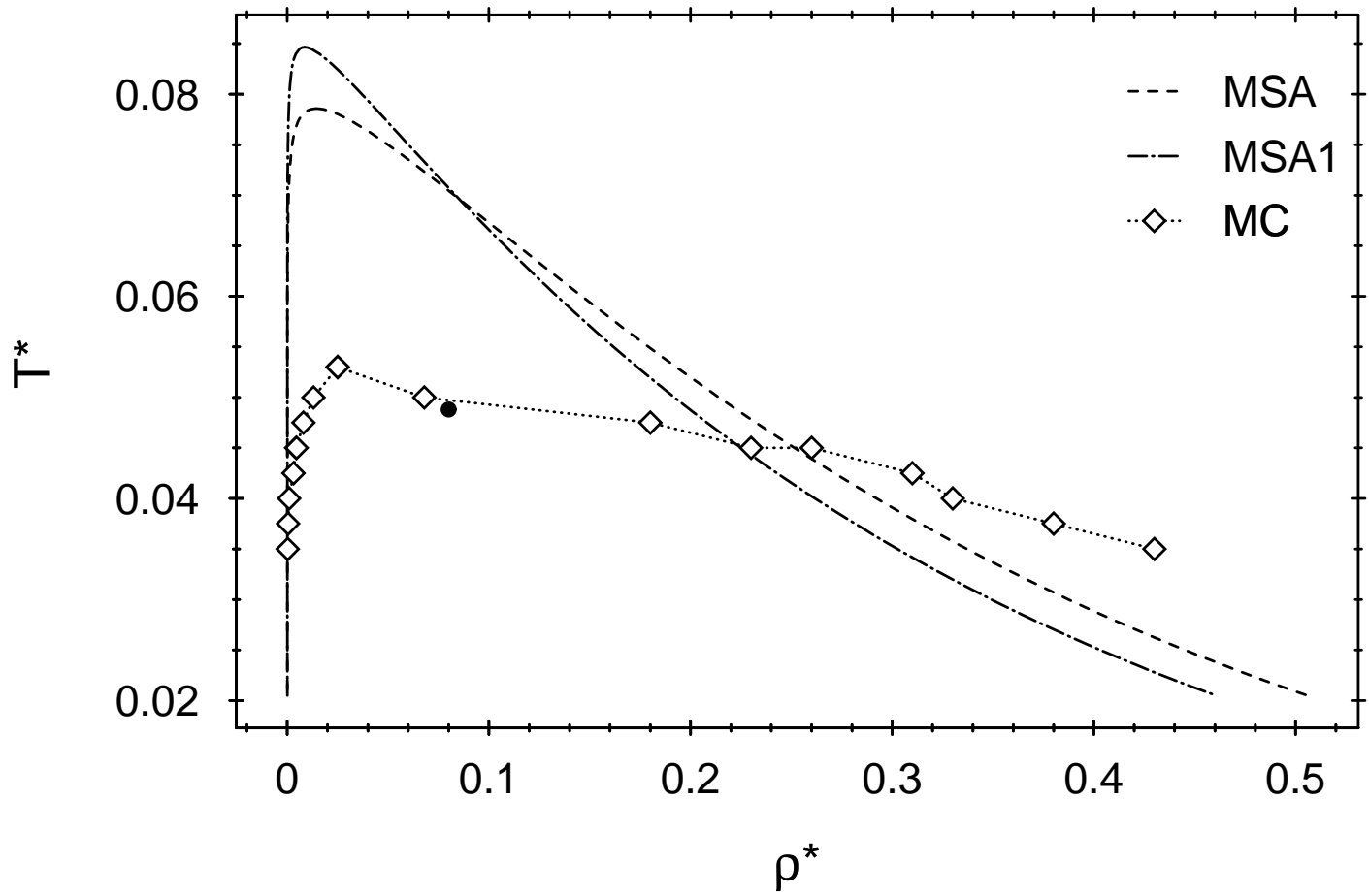
\begin{figure}[htbp]
\caption{The phase diagram of the restricted primitive model in the
  temperature-density plane as given by the approximation schemes MSA
  and MSA1 discussed in the main text. The open symbols denote the results of
  Gibbs ensemble
  Monte Carlo simulations \protect\cite{Pana}. The solid circle
  denotes the most recent simulation estimate of the critical point
  \protect\cite{CaillolRPMMC}. The vapor-liquid-bcc-solid triple point is
  estimated to lie near $T^\ast_t\simeq0.025$ and $\rho^\ast_t\simeq0.5$
  \protect\cite{Smittripel}. The dotted line connecting the Monte-Carlo
  data is a guide to the eye.}
\label{fig:phas}
\end{figure}

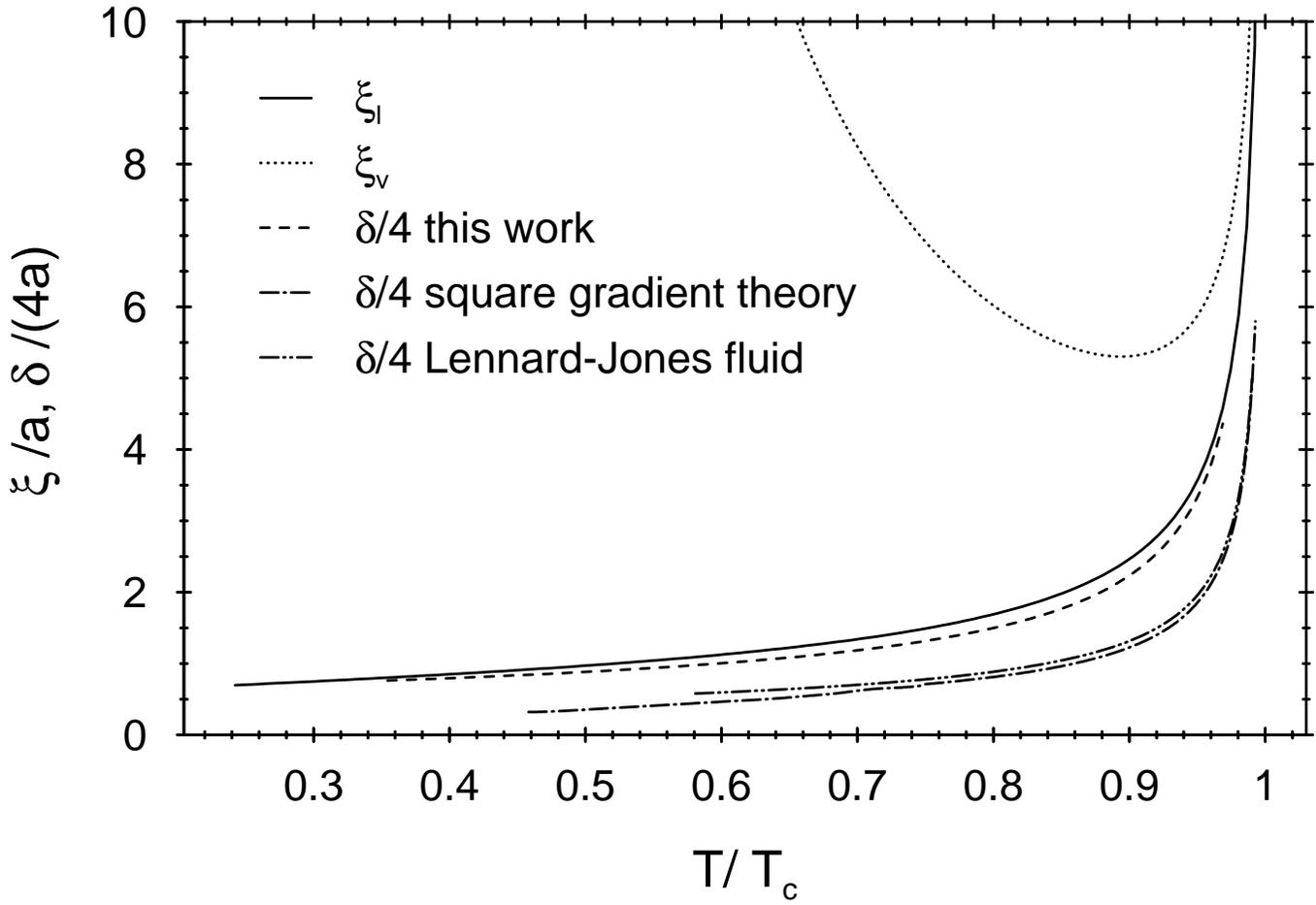
\begin{figure}[htbp]
  \caption{The second moment correlation lengths $\xi_v$ and $\xi_l$
  for density correlations in
  the vapor and liquid phases at two-phase coexistence obtained using the MSA1
  approximation. Also shown is the width $\delta$ (defined by
  Eq.~(\protect\ref{width})) of the liquid-vapor
  interface obtained from the present density-functional theory and from
  the square gradient theory of Ref.~\protect\cite{GT}. All lengths diverge at the
  critical temperature $T_c$. The vapor correlation length increases at low
  temperatures because of the reduced screening in the very dilute
  vapor phase. The interfacial width $\delta$ is very close to $4 \xi_l$ over
  the whole temperature range. At the same reduced temperature the
  interfacial width predicted by the present theory is larger than
  that obtained from the square-gradient theory which in turn is close to the
  result for a
  Lennard-Jones fluid as obtained from a DFT similar to the present one
  \protect\cite{Frodl}.}
  \label{fig:korr}
\end{figure}

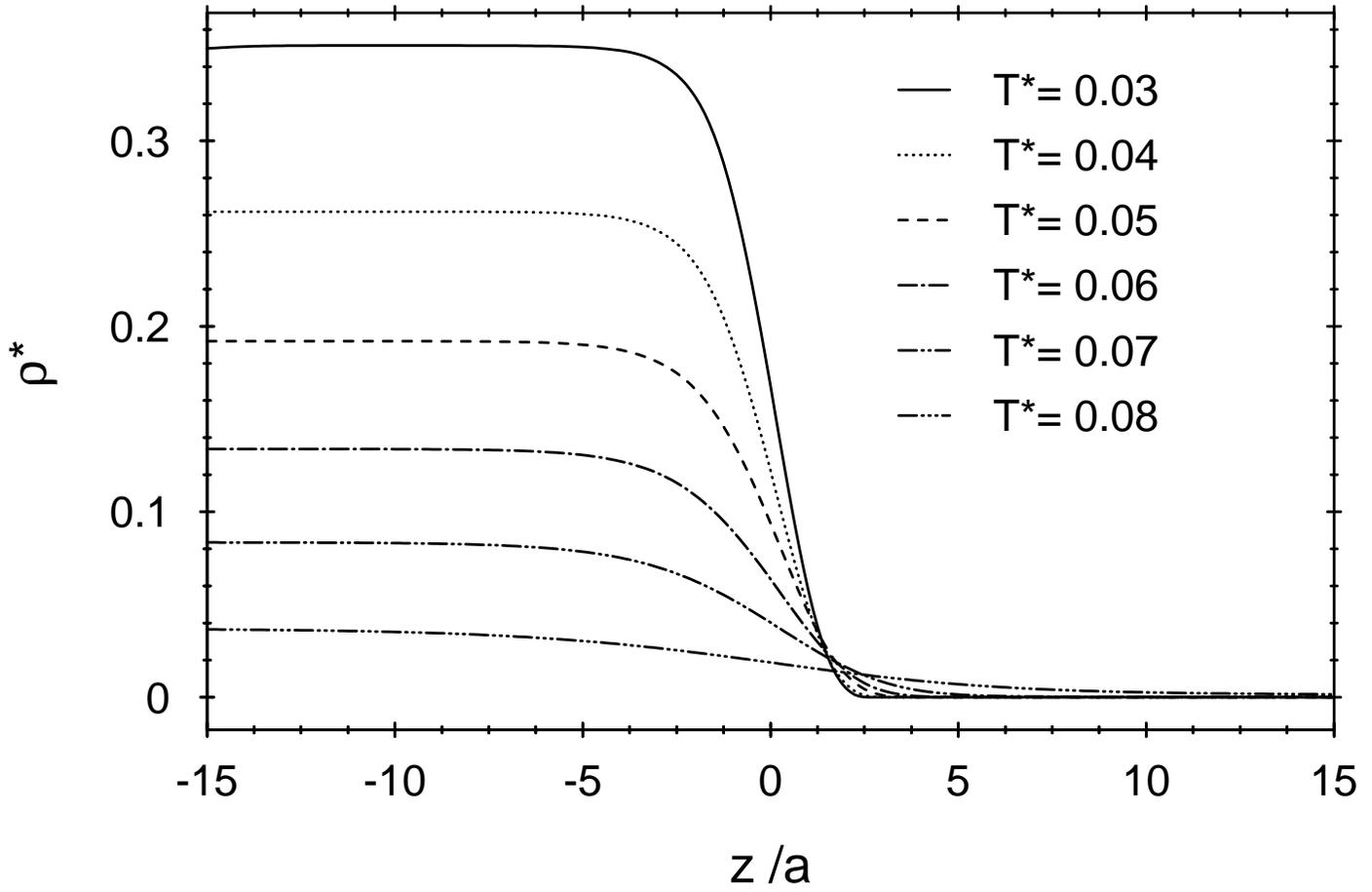
\begin{figure}[htbp]
  \caption{The total density profile of the liquid-vapor interface of the RPM
  at different temperatures,  as obtained from MSA1.}
  \label{fig:profil}
\end{figure}

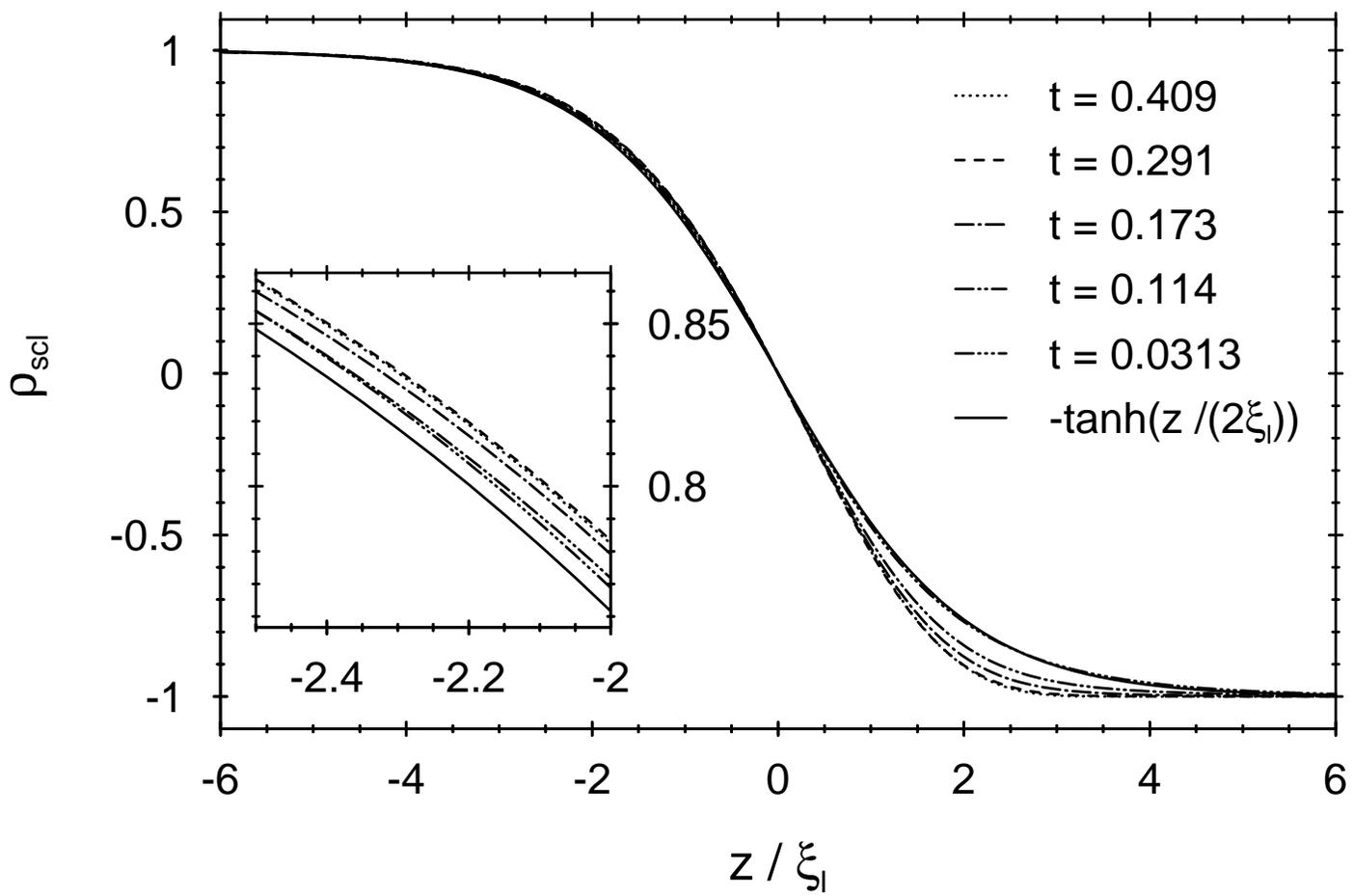
\begin{figure}[htbp]
  \caption{Density profiles, as obtained from MSA1, plotted in scaled form (see
    Eq.~(\protect\ref{rhoscl})) as a function of $z/\xi_l$. The parameter
    $t=1-T/T_c$ measures the deviation from the critical
    temperature. For $t\to 0 $ the scaled profiles approach the
    universal scaling function $-\tanh(z/(2\xi_l))$.}
  \label{fig:rhoscl}
\end{figure}

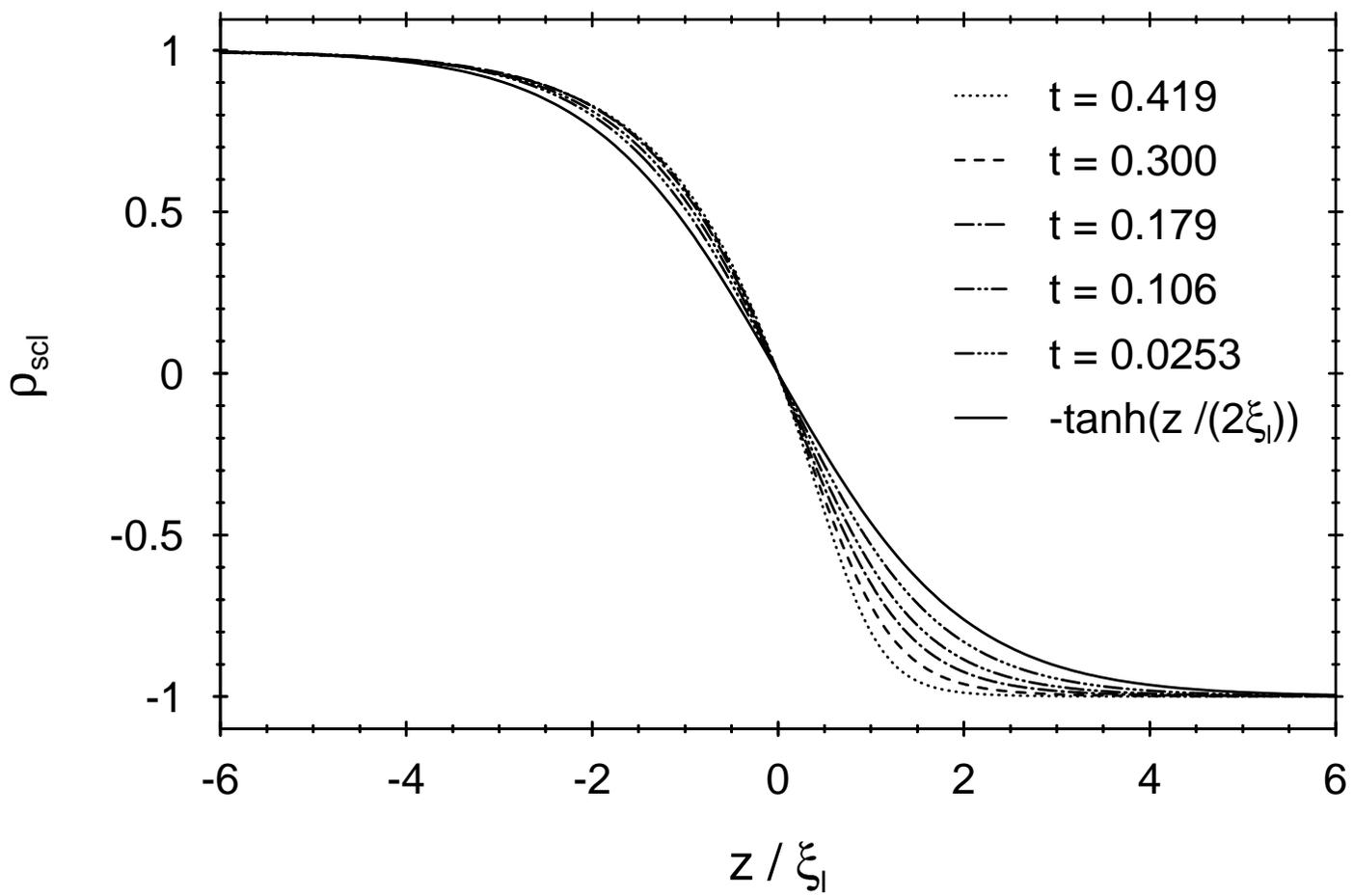
\begin{figure}[htbp]
  \caption{The same plot as Fig.~\protect\ref{fig:rhoscl} but for a Lennard-Jones
    fluid. The results were obtained using the density-functional theory
    of Frodl and Dietrich \protect\cite{Frodl}. Note that
    the approach to the scaling limit is slower and the profiles are
    more asymmetric than for the ionic fluid.}
  \label{fig:rholj}
\end{figure}

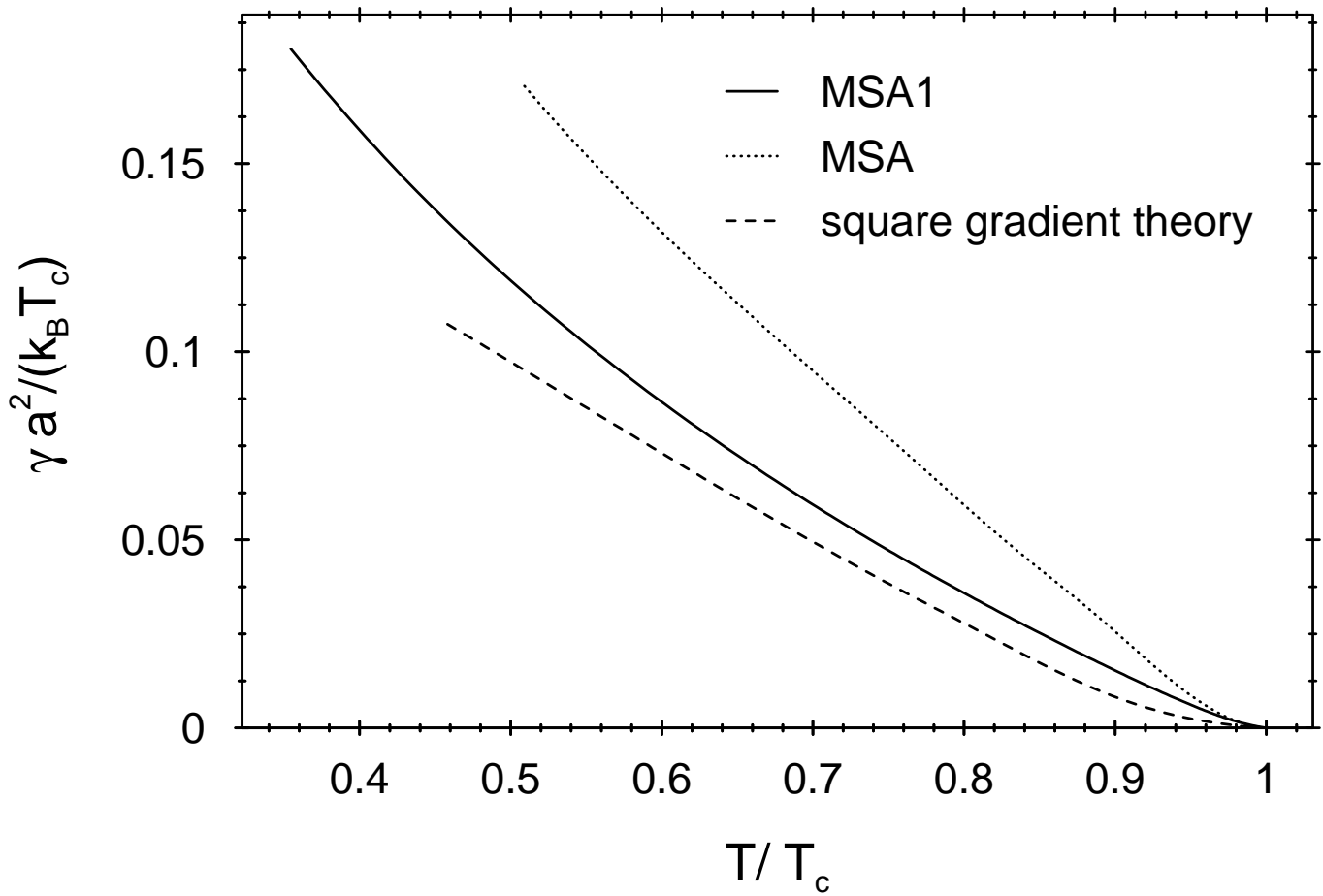
\begin{figure}[htbp]
  \caption{Comparison of the liquid-vapor surface tension $\gamma$ of
    the RPM as given by the present theories MSA and MSA1 and by the square gradient
    theory of Ref.~\protect\cite{GT}. The data have been
    reduced using the critical temperature $T_c$ appropriate to each theory.}
  \label{fig:surf}
\end{figure}

We have developed two, closely related, density-functional theories
for the liquid-vapor interface of the RPM. Both the MSA and the MSA1
are based on approximating the inhomogeneous pair correlation function
by that of a homogeneous bulk liquid at some mean density
$\bar\rho$. The MSA1 involves an additional approximation, i.e., the
integration over the coupling constant (charge) is ignored. The
results which have emerged from this analysis can be summarized
as follows:

\begin{enumerate}
\item MSA and MSA1 yield similar bulk phase diagrams, with the latter
  overestimating the critical temperature $T_c$ and underestimating
  the critical density $\rho_c$ to a further extent than MSA (Fig.~\ref{fig:phas}).

\item The two theories yield similar second moment correlation lengths
  for the density-density correlations in the bulk fluid
  (Fig.~\ref{fig:korr}). They exhibit the correct power law dependences
  in the low-density limit but overestimate their numerical prefactors
  (Eqs.~(\ref{limr0ex})--(\ref{limr0MSA})).
  
\item When suitably scaled to take into account the differences in the
  bulk phase diagram the total density profiles are similar in both
  theories. Both predict that for low temperatures and on intermediate
  length scales the profile approaches its bulk limits faster on the
  vapor side than on the liquid side (Figs.~\ref{fig:profil} and
  \ref{fig:rhoscl}).  This is opposite to what is found in the square-gradient
  theory of Ref.~\cite{GT}. However, the ultimate decay into bulk should be
  determined by the bulk correlation length which is larger on the
  vapor side.

\item The interfacial width $\delta$, defined by Eq.~(\ref{width}), is
  close to four times the {\it liquid} correlation length over the
  whole temperature range. It is larger but has a similar variation with temperature
  as that of a simple liquid (Fig.~\ref{fig:korr}).

\item For $T\to T_c$ the density profiles approach a universal scaling
  form. In contrast to simple fluids this approach is not monotonic as
  a function of temperature (Figs.~\ref{fig:rhoscl} and
  \ref{fig:rholj}).

\item The surface tensions calculated $\gamma$ from the present theories are
   larger than those obtained in Ref.~\cite{GT} and the temperature
  dependence is not as linear (Fig.~\ref{fig:surf}).

\item All theories for the RPM find that for $T/T_c\simeq0.6$ the scaled surface
  tension $\tilde\gamma=\gamma^\ast/T^\ast_c$ is considerably smaller than the
  corresponding ratio for simple atomic liquids. We surmise that this is a
  genuine feature of ionic fluids.

\end{enumerate}

Certain aspects of our theory warrant further discussion. The basis of
our approach is the approximation for the inhomogeneous pair
distribution function $g_{ij}({\bf r},{\bf r}',\{\rho_i({\bf r})\},\alpha)$ which is
the key input for the theory. This has two intriguing consequences for
the asymptotics of the correlation functions generated from the
density functional. The first concerns the low-density behavior of the
second moment correlation length $\xi$, referred to above. That MSA
and MSA1 do not yield the exact limiting behavior for $\xi$ reflects
the fact that even at low densities our approximation for $g_{ij}$
does not capture the proper long wavelength variation, i.e., the
correct coefficient of $k^2$ in the structure factor (Eq.~(\ref{xidef})). 
A possible improvement upon the present approximation for
$g_{ij}({\bf r},{\bf r}',\{\rho_i({\bf r})\},\alpha)$ may be given by
the ``mean density approximation'' (MDA) used in
Refs.~\cite{Schirmacher} and \cite{HendAsh} in which the inhomogeneous pair
distribution function is expanded around its homogeneous limit up to
second order in the deviation $\rho({\bf r})-\rho_b$. But, as for  DFTs
based on a density expansion of the excess free energy, this approach
cannot be applied straightforwardly to the liquid-vapor interface due
to the lack of a unique bulk density that could serve as a starting
point for the expansion.
The second intriguing consequence is that the MSA, through its
integration over $\alpha$ in function space (Eq.~(\ref{DFTexact})), 
yields algebraically decaying bulk
correlation functions and liquid-vapor density profiles whereas the
correct ultimate decay should be exponential in both cases. Once again
this failing must be attributed to the simple approximation employed
for the inhomogeneous pair correlation function. However, it is not
clear  what modifications to the theory should be introduced ---
other than simply omitting the $\alpha$ integration as was done in
MSA1 --- in order to eliminate this feature.

We should also emphasize that the present DFT of the liquid-vapor
interface is a mean-field treatment in that it does not incorporate
the effects of either bulk critical fluctuations or of interfacial
capillary-wave-like fluctuations. In keeping with treatments of simple
atomic fluids we argue that the density profiles and surface
tensions we calculate from our theory are those of the bare or
{\it intrinsic} interface. In order to estimate the additional broadening of the
density profile arising from the interfacial fluctuations one could
perform the standard Gaussian unfreezing of these fluctuations on
the intrinsic interface given by the present theory
\cite{Weeks,EvansLesHouches}. The ``stiffness'' of the interface is
determined by the dimensionless ratio $\omega=k_B
T/(4\pi\gamma\xi^2)$, i.e., the larger the surface tension $\gamma$ the
smaller is the interfacial broadening. Since we predict that in the
RPM the ratio $\tilde\gamma=\gamma a^2/(k_B T)$ is much smaller than
for an atomic fluid at the same reduced temperature $T/T_c$, this
implies that the  broadening due to fluctuations is significantly more
pronounced for the RPM.

There are two other interesting problems to which the present theory
could be applied. The first is the interface between an ionic fluid and a charged
wall. Although this has been investigated using several theoretical
techniques \cite{BlumRev,AttardRev} the advantage of the present
approach is that it incorporates two coexisting phases and thus can
describe wetting phenomena. The second is the liquid-vapor interface of an
ionic fluid in which  cations and anions have unequal diameters. These
more realistic systems have only been tackled within the context of
the gradient expansion \cite{Sluckin}. In both
cases electrical double layers will form giving a local violation of
electroneutrality. However, the bulk correlation functions used as input to the
DFT are only available for neutral systems. Nevertheless, one might
still hope to capture the essential features by employing Eq.~(\ref{rhobar})
for the mapping to the bulk system and by taking account of the non-zero
local charge density only via the Coulomb contribution to the free
energy, Eq.~(\ref{FCoulomb}). A similar assumption has proved
successful for the electrolyte-wall interface in the
density-functional theories of
Refs.~\cite{Mier,Groot,Kierlik,Ghosh}.

\acknowledgements 

We thank B.~G\"otzelmann and R.J.F.~Leote de Carvalho for helpful
discussions. R.E. is grateful for the hospitality of the physics
department of the University of Wuppertal.

\end{document}